\documentclass[aps,prl,twocolumn,amsmath,superscriptaddress]{revtex4-1}  
\usepackage{bbm}
\usepackage{mathrsfs}
\usepackage{amsmath}
\usepackage{amsfonts}
\usepackage[colorlinks=true,citecolor=blue,anchorcolor=blue]{hyperref}
\usepackage{graphicx,epstopdf}
\usepackage{float}
\usepackage{subfigure}
\usepackage{epsfig}
\usepackage{dcolumn}
\usepackage{bm}
\usepackage{color}
\usepackage{natbib}
\usepackage{amssymb}
\usepackage{xcolor}
\usepackage{braket}
\usepackage{hyperref}
\usepackage{url}
\begin{document}
\title{Quenching of Nonrelativistic $p$-Wave Spin Splitting by Reduced $c\text{-}f$ Coupling in $\text{CeNiAsO}$}

\author{Xinnuo Zhang}
\thanks{Equal contributions}
\affiliation{National Synchrotron Radiation Laboratory and School of Nuclear Science and Technology, University of Science and Technology of China, Hefei, 230026, China}

\author{Zhicheng Jiang}
\thanks{Equal contributions}
\affiliation{National Synchrotron Radiation Laboratory and School of Nuclear Science and Technology, University of Science and Technology of China, Hefei, 230026, China}

\author{Shibo Shen}
\thanks{Equal contributions}
\affiliation{School of Emerging Technology, University of Science and Technology of China, Hefei 230026, China}

\author{Jian Yuan}
\thanks{Equal contributions}
\affiliation{State Key Laboratory of Quantum Functional Materials, School of Physical Science and Technology, ShanghaiTech University, Shanghai 201210, China}

\author{Junseo Yoo}
\affiliation{Department of Physics and Astronomy, Seoul National University, Seoul 08826, Republic of Korea}

\author{Xun Ma}
\affiliation{National Synchrotron Radiation Laboratory and School of Nuclear Science and Technology, University of Science and Technology of China, Hefei, 230026, China}

\author{Mao Ye}
\affiliation{Shanghai Synchrotron Radiation Facility, Shanghai Advanced Research Institute,
Chinese Academy of Sciences, Shanghai 201210, China}
\affiliation{National Key Laboratory of Materials for Integrated Circuits, Shanghai Institute of Microsystem and Information Technology, Chinese Academy of Sciences, Shanghai 200050, China}

\author{Jishan Liu}
\affiliation{Shanghai Synchrotron Radiation Facility, Shanghai Advanced Research Institute,
Chinese Academy of Sciences, Shanghai 201210, China}
\affiliation{National Key Laboratory of Materials for Integrated Circuits, Shanghai Institute of Microsystem and Information Technology, Chinese Academy of Sciences, Shanghai 200050, China}

\author{Zhengtai Liu}
\email{liuzt@sari.ac.cn}
\affiliation{Shanghai Synchrotron Radiation Facility, Shanghai Advanced Research Institute,
Chinese Academy of Sciences, Shanghai 201210, China}
\affiliation{National Key Laboratory of Materials for Integrated Circuits, Shanghai Institute of Microsystem and Information Technology, Chinese Academy of Sciences, Shanghai 200050, China}

\author{Changyoung Kim}
\affiliation{Department of Physics and Astronomy, Seoul National University, Seoul 08826, Republic of Korea}

\author{Yanfeng Guo}
\email{guoyf@shanghaitech.edu.cn}
\affiliation{State Key Laboratory of Quantum Functional Materials, School of Physical Science and Technology, ShanghaiTech University, Shanghai 201210, China}
\affiliation{ShanghaiTech Laboratory for Topological Physics, ShanghaiTech University, Shanghai 201210, China}

\author{Yilin Wang}
\email{yilinwang@ustc.edu.cn}
\affiliation{School of Emerging Technology, University of Science and Technology of China, Hefei 230026, China}
\affiliation{Hefei National Laboratory, Hefei 230088, China}

\author{Dawei Shen}
\email{dwshen@ustc.edu.cn}
\affiliation{National Synchrotron Radiation Laboratory and School of Nuclear Science and Technology, University of Science and Technology of China, Hefei, 230026, China}

\begin{abstract}

The application of spin-space group symmetries to noncollinear antiferromagnets has led to the prediction of odd-parity, nonrelativistic spin splittings, making the physical realization of a practical $p$-wave magnet a central pursuit in spintronics. The layered heavy-fermion oxypnictide $\text{CeNiAsO}$ has been widely regarded as a prototypical platform to verify this paradigm. Here, we investigate the electronic structure of single-crystal $\text{CeNiAsO}$ using high-resolution, ultra-low-temperature and resonant angle-resolved photoemission spectroscopy (ARPES), and $ab-initio$ calculations. Across the consecutive magnetic transitions into the ordered phases, our spectroscopic data reveal neither the expected band folding associated with a spin density wave nor any observable $p$-wave spin splitting, demonstrating that the conduction bands retain full degeneracy. By tracking the temperature dependence of the Ce 4$f$ spectral weight via resonant ARPES, we find negligible $c\text{-}f$ hybridization near the Fermi level within magnetically ordered states, confirming that the Ce 4$f$ electrons reside close to the localized limit. Our findings establish a clear many-body constraint on the projection of real-space magnetic symmetries onto momentum-space electronic bands, demonstrating that symmetry classifications constitute a necessary framework but are not a sufficient condition for nonrelativistic spin splittings in the presence of strong electronic correlations.
\end{abstract}

\maketitle

\begin{figure}[htbp]
\centering
\includegraphics[width=\linewidth]{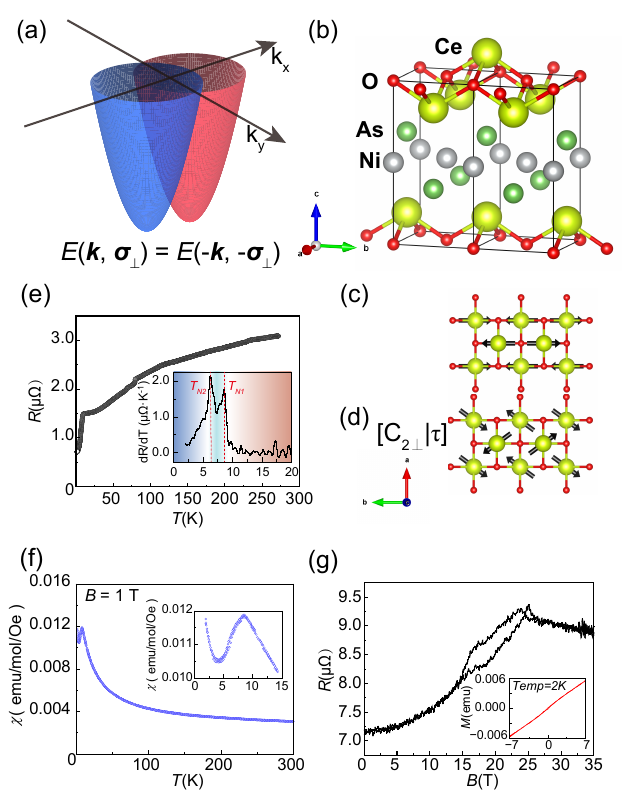}
\caption{Basic properties and transport measurements.
(a) Spin-polarized Fermi surfaces of a $p$-wave magnet with the dispersion satisfying $E(\mathbf{k}, \sigma_\perp)=E(-\mathbf{k},-\sigma_\perp)$. 
(b) Tetragonal crystal structure of $\text{CeNiAsO}$ showing alternating stacked layers. 
(c) Incommensurate SDW below $T_{\mathrm{N1}}$: a magnetic cycloid approximately treated as a longitudinal spin-density wave. 
(d) Commensurate coplanar order below $T_{\mathrm{N2}}$: a coplanar helical spin structure. 
(e) Resistivity $R(T)$ of $\text{CeNiAsO}$; inset shows $dR/dT$ to resolve the two-step transitions, where $T_{N1}$=8.66K, $T_{N2}$=6.23K. 
(f) Susceptibility $\chi(T)$ measured at $B=1$~T along the $c$ axis; inset magnifies the low-$T$ region. 
(g) Magnetoresistance $R(B)$ up to 35~T at 2~K temperature; inset shows $M(B)$ from -7~T to 7~T.
}
\label{fig1}
\end{figure}

The realization of nonrelativistic spin splittings independent of spin-orbit coupling (SOC) has emerged as a central frontier in quantum magnetism and spintronics~\cite{Smejkal2020Crystal, Smejkal2022Beyond, Yuan2020Giant, Smejkal2022Emerging, song2025altermagnets, jungwirth2026symmetry}. This paradigm was initiated by altermagnets, where collinear antiferromagnetic (AFM) structures unlock an even-parity ($d$- or $g$-wave) spin splitting at the electron-volt scale for high-efficiency spin currents~\cite{song2025altermagnets, Zhang2026ARPES, Liu2026Symmetry, Roig2024Minimal, samanta2025spin, fu2025all, zhang2025electrical, noh2025tunneling, liu2024giant,jiang2025metallic,zhang2025crystal}. Stimulated by this, spin-space group (SSG)~\cite{Liu2022SpinGroup, Xiao2024SpinSpace, chen2025unconventional, liu2026symmetryclassidfication, chen2024enumeration, jiang2024enumeration, brinkman1966theory} based classifications have generalized this phenomenon to noncollinear (NCL) magnetic structures that break spatial inversion symmetry while preserving a generalized time-reversal symmetry, predicting odd-parity $p$-wave magnets~\cite{Priessnitz2026Ferroelectric, Leon2025Hybrid, zhu2026floquet, Rickelt2026Hatsugai, Lin2025OddParity,zhuang2025odd, liu2026light, zhuang2025odd, dsouza2026odd, luo2025spin}. 
It shifts the spin-polarized Fermi surfaces asymmetrically in momentum space, satisfying $E(\mathbf{k}, \sigma_\perp) = E(-\mathbf{k}, -\sigma_\perp)$ and generating a unique $p$-wave spin texture with opposite polarizations on opposing sides of the Fermi surface [Fig.~\ref{fig1}(a)]. Acting as a nonrelativistic counterpart to relativistic Rashba/Dresselhaus SOC, these profiles are governed by strong exchange energy scales and highly attractive for nonrelativistic Edelstein responses and for engineering unconventional triplet pairing toward topological superconductivity~\cite{Chakraborty2025Highly, Zhao2025Altermagnetism, Sukhachov2025Coexistence, Ezawa2025Out, Brekke2024Minimal, yamada2025metallic, sun2025pseudo, soori2025crossed, ezawa2024topological, nagae2025flat, kokkeler2025quantum, Song2025Electrical, wang2026microscopic, lee2026incommensuration, zeng2025tunneling, luo2026hidden}.

The heavy-fermion oxypnictide $\text{CeNiAsO}$ has been proposed as a promising platform to host $p$-wave spin splitting driven by its coplanar NCL AFM order on $\text{Ce}$-sites~\cite{Hellenes2023P, Yu2025OddParity, Zhao2025Altermagnetism, error, Chakraborty2025Highly, Sukhachov2025Coexistence, Ezawa2025Out, zhou2025anisotropic, Priessnitz2026Ferroelectric, luo2025spin, Wu2019Incommensurate, Lu202375As}. To date, however, a direct momentum-resolved spectroscopic verification of such odd-parity spin splitting remains entirely elusive across all candidate materials. 
Although SSG symmetry classifications establish the necessary condition for odd-parity magnets, the spin-splitting in actual materials strongly depends on the assumed well-defined single-particle band structures in momentum space. 
In strongly correlated heavy-fermion $f$-electron materials, this single-particle framework must confront the intrinsic many-body competition between the local-moment behavior of $f$-electrons and the phase-coherent hybridization with the conduction $c$-electrons. To enable the momentum-space spin splitting, the correlated $4f$ electrons must either form coherent, dispersive bands that split directly, or develop a sufficiently strong $c\text{-}f$ exchange coupling to transfer the odd-parity magnetic potentials to the itinerant conduction bands. Both scenarios dictate a phase-coherent $c\text{-}f$ entanglement, leaving the validity of static symmetry predictions in the presence of strong electronic correlations an open but
crucial question.

In this Letter, we address this issue by tracking the low-energy electronic structure of single-crystal $\text{CeNiAsO}$ using high-resolution angle-resolved photoemission spectroscopy (ARPES). Across the magnetic transitions into the coplanar NCL AFM state, our momentum-resolved spectra reveal strict, unlifted electronic degeneracy with no detectable $p$-wave band splitting at zero magnetic field. Furthermore, resonant ARPES (rARPES) measurements detect no spectral signatures of coherent $\text{Ce}$ $4f$ hybridization with the conduction bands near the Fermi level ($E_{\text{F}}$) within the magnetically ordered state. This strongly suppressed hybridization confirms that the $f$ electrons remain nearly fully localized, thereby isolating the conduction electrons from the expected odd-parity magnetic background such that there is no spin splitting. Our results demonstrate that while SSG  provides a necessary framework for identifying the symmetry of altermagnets and odd-parity magnets, it is not a sufficient condition for the expected spin splitting in the systems with strong electronic correlations.

Details of our experiments and $ab-initio$ calculations can be found in Section 1 of the Supplemental Material (SM). $\text{CeNiAsO}$ crystallizes in a tetragonal $\text{ZrCuSiAs}$-type structure with space group $P4/nmm$ (No. 129) [Fig.~\ref{fig1}(b)]~\cite{blanchard2010effects,luo2011ceniaso,luo2014heavy}. Prior neutron scattering and transport studies resolve two successive magnetic transitions upon cooling from the paramagnetic state: a spin-density-wave (SDW) phase below $T_{\text{N1}} \approx 9$~K [Fig.~\ref{fig1}(c)], and then a coplanar NCL AFM state below $T_{\text{N2}} \approx 6$~K [Fig.~\ref{fig1}(d)]~\cite{Wu2019Incommensurate,zhou2025anisotropic}. Governed by the non-trivial SSG operation $[C_{2\perp}|\mathbf{t}]$, where $\mathbf{t}$ is a fractional spatial translation, this state was predicted to exhibit an asymmetric $p$-wave spin splitting [$E(\mathbf{k}, \sigma_\perp) = E(-\mathbf{k}, -\sigma_\perp)$] with nonzero spin textures along the $C_{2\perp}$ rotation axis [Fig.~\ref{fig1}(a)]. 

\begin{figure}[htbp]
\centering
\includegraphics[width=9cm]{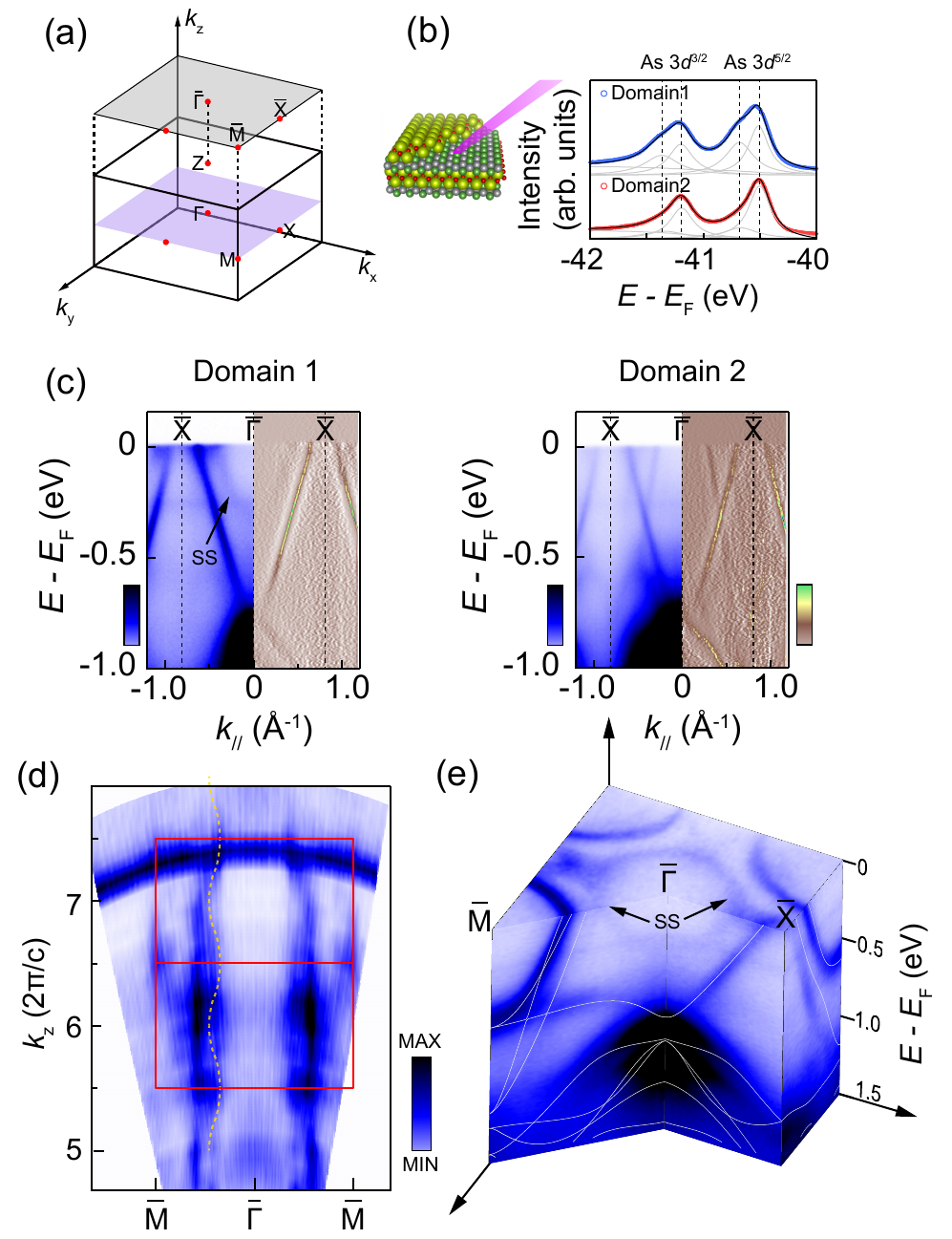}
\caption{Paramagnetic-state ARPES results.
(a) 3D Brillouin zone of $\text{CeNiAsO}$ and its projection onto the (001) surface Brillouin zone. 
(b) Cleaved $\text{CeNiAsO}$ surface with possible terminations; spatially resolved As 3$d$ core-level spectra for two surface domains (gray curves: Lorentzian fits). 
(c) ARPES intensity and second-derivative images along $\overline{\Gamma}$-$\overline{X}$ for the two domains; arrow: an extra surface-related band in domain 1. 
(d) Photon-energy-dependent ARPES intensity along $\overline{M}$-$\overline{\Gamma}$-$\overline{M}$ as a function of $k_z$; dashed line guides the eye for periodic $k_z$ dispersion, the inner potential is 10 eV . 
(e) 3D ARPES intensity on the $\overline\Gamma$-$\overline{X}$-$\overline{M}$ plane, including the Fermi-surface map and band dispersions; white curves: renormalized bulk-band calculations; black arrows: additional surface states.}
\label{fig2}
\end{figure}

To benchmark our samples against established structural and magnetic properties, single-crystal X-ray diffraction (XRD) was performed [Fig.~S1, TABLE I and II in SM], revealing sharp Bragg reflections with no detectable impurity phases. The temperature-dependent resistivity $R(T)$ [Fig.~\ref{fig1}(e)] exhibits clear anomalies, with the derivative $dR/dT$ resolving magnetic transitions at 8.66~K and 6.23~K, respectively, in good agreement with literature~\cite{Wu2019Incommensurate,zhou2025anisotropic}. The magnetic susceptibility $\chi(T)$ [Fig.~\ref{fig1}(f)], measured under $B = 1$~T along the $c$-axis, displays a pronounced peak at $T_{\text{N1}}$ and a subtle shoulder around $T_{\text{N2}}$ under both zero-field/field cooling conditions, substantiating the two-step magnetic ordering. Moreover, high-field magnetotransport measurements reveal a hysteretic magnetoresistance between 15 and 25~T [Fig.~\ref{fig1}(g)], indicative of magnetic frustration within the ordered Ce layers. As shown in the inset of Fig.~\ref{fig1}(g), the magnetization increases monotonically up to 7~T, while the negative magnetoresistance emerging beyond $B > 25$~T indicates that the system approaches a field-induced polarized magnetic state.

\begin{figure*}[htbp]
\centering
\includegraphics[width=17.5 cm]{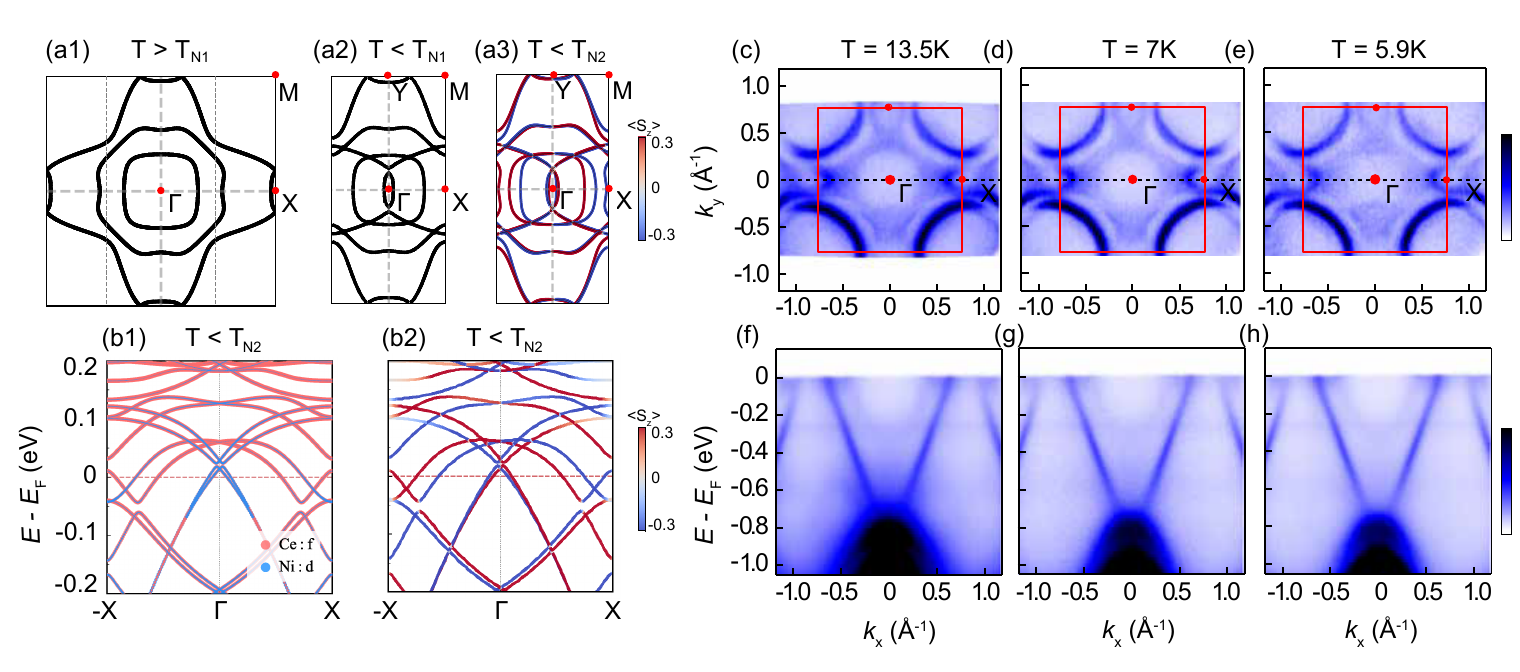}
\caption{Evolution of electronic structures as decreasing temperatures. 
(a1)-(a3) The DFT calculated Fermi surfaces for the nonmagnetic phase, SDW phase below $T_{\mathrm{N1}}$, and $p$-wave magnetic phase below $T_{\mathrm{N2}}$, respectively. The magnetic order doubles the unit-cell and folds the bands along $\Gamma$-X direction, and shows a $p$-wave spin splitting with nonzero $s_z(\mathbf{k})$. The incommensurate SDW phase below $T_{\mathrm{N1}}$ is approximated as a $2\times1$ collinear AFM order in calculation.  
(b1) The bands are projected onto Ce-$f$ (pink) and Ni-$d$ (blue) orbitals. (b2) Spin projections of $s_z(\mathbf{k})$. 
(c)-(e) ARPES Fermi-surface maps at 13.5~K, 7~K, and 5.9~K; red dots: high-symmetry points; red squares: appended Brillouin zone. 
(f)-(h) ARPES spectra along $\Gamma$--X at the same temperatures corresponding to (c)-(e).}
\label{fig3}
\end{figure*}

The bulk Brillouin zone (BZ) of $\text{CeNiAsO}$ and its projection onto the (001) surface BZ are schematically illustrated in Fig.~\ref{fig2}(a). Because $\text{CeNiAsO}$ naturally cleaves along polar planes to yield either $[\text{CeO}]^{+}$- or $[\text{NiAs}]^{-}$-terminated surfaces, we characterized these cleavage profiles using micro-focused ARPES. Figure~\ref{fig2}(b) displays representative $\text{As}$ $3d$ core-level spectra from two distinct domains. Specifically, domain 1 exhibits a pronounced satellite splitting within the $\text{As}$ $3d^{3/2}$ and $3d^{5/2}$ spin-orbit doublets, whereas this feature is substantially suppressed in domain 2. This distinct line-shape disparity allows us to assign domain 1 to the $[\text{NiAs}]^{-}$-terminated surface layer. Paramagnetic state ARPES cuts along $\overline{\Gamma}\text{-}\overline{X}$ [Fig.~\ref{fig2}(c), with second-order derivatives on the right halves] confirm that both domains possess highly similar bulk bands. However, because domain 2 hosts additional folding replicas driven by cleavage-induced surface reconstructions, we focus our subsequent systematic analysis on the domain 1 (datasets of domain 2 are compiled in Fig.~S3).

By mapping the $k_z\text{-}k_{\parallel}$ photoemission intensity dispersion on domain 1 [Fig.~\ref{fig2}(d)], we resolved a clear periodic modulation along the $k_z$ axis that matches density functional theory (DFT) bulk band calculations (dashed lines and Fig.~S7 and S8), establishing $h\nu = 90$~eV as the bulk $\Gamma\text{-}X\text{-}M$ plane. Figure~\ref{fig2}(e) summarizes the paramagnetic electronic structure mapped within the high-symmetry plane, displaying the Fermi surface alongside the corresponding dispersions. The overlaid DFT calculations, treating the $\text{Ce}$ $4f$ electrons as localized core states, reproduce the primary experimental bulk dispersions exceptionally well under a modest renormalization factor of $\sim 0.7$, while a few mismatched spectral features (marked by black arrows) can be assigned to surface states.

Having established the paramagnetic state, we next tracked the low-energy bands across the successive magnetic phase transitions of $\text{CeNiAsO}$. As illustrated in Figs.~\ref{fig3}(a1)--(a3) and (b1)--(b2), DFT calculations without considering the strong electron correlations of Ce-$4f$ electrons, i.e., in the itinerant limit, predict two distinct electronic signatures upon entering the ordered states~\cite{Hellenes2023P, Yu2025OddParity}. Firstly, the approximate $2\times1$ collinear SDW phase below $T_{\text{N1}}$ doubles the magnetic unit cell, reconstructing the BZ and folding the Fermi surface along the $\Gamma$-$X$ direction [Fig.~\ref{fig3}(a2)]. Second, the onset of the coplanar NCL AFM order below $T_{\text{N2}}$ is predicted to exhibit sizeable $p$-wave spin splitting, in particular, along the $\Gamma$-$X$ direction [Figs.~\ref{fig3}(a3) and (b2)]. As shown in Figs.~\ref{fig3}(b1) and (b2), the calculated $4f$ bands near $E_{\text{F}}$ are highly dispersive and strongly hybridized with the $\text{Ni}$-$3d$ orbitals. Consequently, the predicted spin splittings are jointly sustained by both the itinerant $\text{Ce}$-$4f$ and $\text{Ni}$-$3d$ electrons. This itinerant picture is consistent with previous theoretical reports~\cite{Hellenes2023P,Chakraborty2025Highly,error,Yu2025OddParity}.

Remarkably, our experimental observations stand in sharp contrast to these theoretical predictions. Figures~\ref{fig3}(c)--(e) display the Fermi surfaces systematically mapped across the three distinct magnetic regimes at 13.5~K, 7~K, and 5.9~K, respectively. The global Fermi surface topography remains entirely invariant, exhibiting no signatures of folded replicas or split bands. This persistent rigidity is further verified by the high-resolution $\Gamma\text{-}X$ energy-momentum dispersions [Fig.~\ref{fig3}(f)--(h)], in which the primary electron pocket evolves smoothly across both $T_{\text{N1}}$ and $T_{\text{N2}}$ without any detectable back-folding or symmetry-breaking reconstruction. Most importantly, the low-temperature bands below $T_{\text{N2}}$ retain strict spin degeneracy, displaying no evidence of the predicted $p$-wave nonrelativistic splitting as illustrated in Fig.~\ref{fig3}(b2). This complete quenching of the expected spin splitting is further substantiated by ultra-low-temperature laser-ARPES scans down to 2.9~K (Fig.~S4) and systematic temperature-dependent surveys extending across multiple distinct $k_z$ planes (Fig.~S2, S5 and S6), both of which consistently resolve a single band feature devoid of any energy or momentum splitting.

\begin{figure*}[htbp]
\centering
\includegraphics[width=17cm]{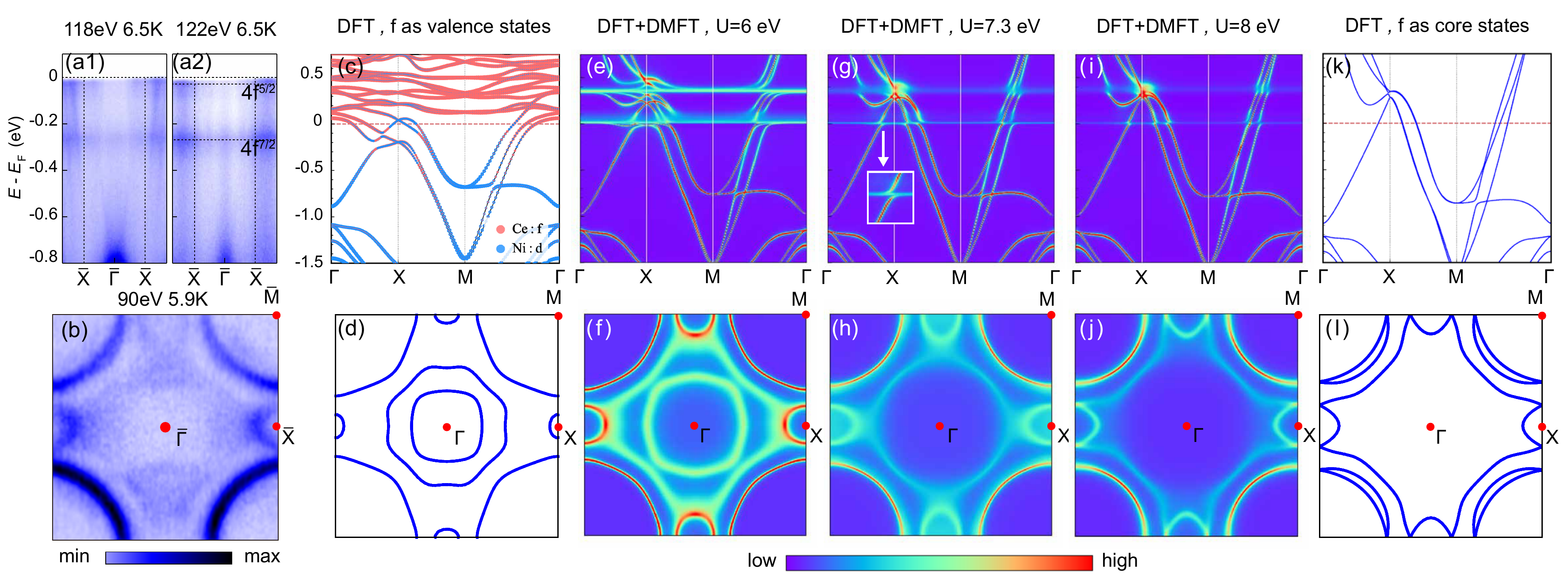}
\caption{Reduced c–f coupling due to localized f electrons.
(a1),(a2) ARPES intensity along $\Gamma$--X at 6.5 K with near-resonant photon energies 118 eV and 122 eV, respectively. Dashed lines: first Brillouin-zone boundaries; solid line: $E_F$. In (a2), horizontal dashed lines mark resonant $4f$ levels, $4f^{5/2}$ and $4f^{7/2}$. (b) ARPES Fermi surface at 5.9 K; black arrow indicates the surface state (SS).
(c),(d) DFT band structure and Fermi surface in nonmagnetic state with Ce-$f$ as valence states. 
(e)-(j) DFT+DMFT results in paramagnetic state for three values of Hubbard $U$. 
(k),(l) DFT results in nonmagnetic state with Ce-$f$ as core states.  SOC is included in calculations.
}
\label{fig4}
\end{figure*}

This unexpected decoupling between the magnetic order in real space and the spin splitting in momentum space points to a deeply localized regime where the $c\text{-}f$ coupling is strongly suppressed, effectively screening the conduction electrons from the underlying magnetic background. To elucidate its microscopic origin, we performed rARPES measurements across the $\text{Ce}$ $4d\text{-}4f$ threshold ($h\nu = 118$--$122$~eV) to systematically enhance the 4$f$ spectral weight along $\Gamma\text{-}X$ [Figs.~\ref{fig4}(a1) and (a2)]. This allows us to discern whether the missing spin splitting stems from a fundamental lack of phase-coherent $c\text{-}f$ hybridization, or merely from a suppressed photoemission cross-section. Despite the massive resonance-induced enhancement of $4f$ sensitivity, particularly at the 122~eV maximum, the spectra resolve two sharply localized 4$f$ flat bands traversing the pre-existing nonmagnetic bands, with no discernible signatures of coherent $c\text{-}f$ hybridization or $f$ band splitting near $E_{\text{F}}$. These rARPES results unambiguously establish that the absence of nonrelativistic spin splitting is not an artifact of insufficient photoemission intensity. Instead, the low-energy physics remains dominated by deeply localized $4f$ moments that fail to form phase-coherent itinerant quasiparticle bands within the magnetically ordered ground state.

To provide a microscopic basis for this localized ground state, we then performed systematic DFT plus dynamical mean-field theory ($\text{DFT+DMFT}$) calculations in the paramagnetic phase to track the evolution of the $\text{Ce}$ $4f$ states as a function of the local Coulomb repulsion $U$. As increasing $U$ from 6 eV to 8 eV, the Ce-$4f$ bands significantly deviate from the itinerant limit [Figs.~\ref{fig4}(c)-(d), DFT calculation with Ce-$4f$ as valence states] and become highly incoherent atomic-like levels pinned near $E_{\text{F}}$, and the $c\text{-}f$ hybridization is significantly suppressed [Figs.~\ref{fig4}(e)-(j)], which asymptotically converges toward the fully localized limit where the $4f$ electrons are frozen as core states in $\text{DFT}$ calculations [Figs.~\ref{fig4}(k)-(l)]. This evolution directly substantiates a comprehensive $c\text{-}f$ orbital decoupling. We note that the calculated Fermi surface at $U = 7.3$~eV in the paramagnetic state [Fig.~\ref{fig4}(h)] exhibits an exceptional match with our experimental data in the magnetically ordered state [Fig.~\ref{fig4}(b)]. While a trace of residual $c\text{-}f$ hybridization remains, as shown in the inset of Fig.~\ref{fig4}(g), its dynamic scattering amplitude is heavily suppressed. This confirms that the $\text{Ce}$-$4f$ electrons reside deep within the localized-moment limit. 
This underscores a correlation-enforced boundary where the localized $4f$ electrons fail to transfer their magnetism to the conduction electrons to induce a spin splitting of the Fermi surface.

The failure of the conduction electrons to inherit the magnetism directly reflects a deep positioning of $\text{CeNiAsO}$ within the classic Doniach phase diagram, which scales the microscopic balance between intersite RKKY interactions ($T_{\text{RKKY}} \propto J_{c\text{-}f}^2$) and intrasite Kondo screening ($T^* \propto \exp(-1/\rho J_{c\text{-}f})$)~\cite{doniach1977kondo}. In the  predictions of the $p$-wave magnets based on SSG symmetries and DFT calculations, it was implicitly assumed that the candidate materials reside on the right side of the Doniach framework, developing a heavy-fermion coherence temperature ($T^*$) well above the magnetic ordering temperature ($T_{\text{N}}$)~\cite{Kirchner2020Colloquium, doniach1977kondo, stewart1984heavy, coleman2006heavy, Hellenes2023P, Yu2025OddParity, Zhao2025Altermagnetism, error}. In that scenario, pre-established coherent $c\text{-}f$ hybridization would integrate the $\text{Ce}$ $4f$ electrons into the Fermi sea, allowing the low-temperature odd-parity exchange field to easily transfer to the itinerant heavy quasiparticles and lift the Kramers degeneracy. However, $\text{CeNiAsO}$ actually occupies the opposite, localized extreme of the Doniach continuum, in which weak $c\text{-}f$ coupling locks the system into an RKKY-dominated regime ($T_{\text{RKKY}} \gg T^*$)~\cite{Kirchner2020Colloquium, doniach1977kondo, stewart1984heavy, coleman2006heavy, paschen2021quantum, patil2016arpes, kummer2015temperature, jw2005kondo, ernst2011emerging, burdin2000coherence, coleman2015introduction}. Our spectroscopic and theoretical results confirm that these $\text{Ce}$ $4f$ states remain entirely confined to a highly incoherent atomic-like levels, leaving the conduction bands completely detached from the magnetic skeleton in terms of low-energy quantum coherence. 

This creates a severe paradox: while the conduction electrons act as virtual high-energy intermediaries to mediate the RKKY exchange and stabilize the long-range magnetic ordering, they remain dynamically decoupled from the ordered $4f$ moments near $E_{\text{F}}$ due to the lack of phase-coherent $c\text{-}f$ entanglement. Consequently, the low-energy exchange potential experienced by conduction bands remains symmetric, suppressing the nonrelativistic spin splitting and preserving full degeneracy.
This $c\text{-}f$ decoupling reveals a critical boundary when applying the SSG-based symmetry classifications to the strongly correlated materials. While the real space magnetic ordering provides the structural symmetry required by theory, it is not a sufficient condition to realize the desired spin splitting of electronic bands in the momentum space. Our findings suggest that future materials searching for altermagnets or odd-parity magnets in $f$-electron systems need to evaluate the competition between $T_{\text{RKKY}}$ and $T^*$ alongside the structural and magnetic ordering classifications, preferentially targeting systems with intrinsic carrier itineracy or robust Kondo lattices with higher coherence temperatures.

In summary, our detailed ARPES investigation of single-crystal $\text{CeNiAsO}$ provides a definitive arbitration on the physical realization of odd-parity nonrelativistic spin splittings. Across the consecutive magnetic transitions, the conduction bands retain full Kramers degeneracy, exhibiting neither the expected $p$-wave spin splitting nor SDW-induced unidirectional band folding, which directly challenges the symmetry-based predictions. This absence of spin splitting stems from the full localization of $\text{Ce}$ $4f$ electrons and the lack of coherent $c\text{-}f$ hybridization, 
thereby quenching the exchange potential of the conduction electrons. Our findings demonstrate that symmetry classifications constitute a necessary framework but not a sufficient condition for nonrelativistic spin splittings, highlighting a rigid many-body constraint.

\textit{Acknowledgment}-This work is supported by National Key R\&D Program of China (Grants No. 2023YFA1406304, 2023YFA1406100 and 2024YFA1408103), National Natural Science Foundation of China (Grant No. 12494593), HFNL Self-Deployed Project (Grant No. ZB2602000302), Anhui Provincial Natural Science Foundation (Grant No. 2408085J003), and the New Cornerstone Science Foundation. Y. F. Guo acknowledges the Analytical Instrumentation Center (SPST-AIC10112914) and the Double First-Class Initiative Fund of ShanghaiTech University. Y.L. Wang was supported by Quantum Science and Technology-National Science and Technology Major Project (No. 2021ZD0302803). Z.C.J. acknowledges the China National Postdoctoral Program for Innovative Talents (BX20240348) and Xiaomi Young Talents Program. We thank the Shanghai Synchrotron Radiation Facility (SSRF)
of BL03U (31124.02.SSRF.BL03U) for assistance on the ARPES measurements.

\textit{Data availability}-The data that support the findings of this article are not publicly available. The data are available from the authors upon reasonable request.

\bibliographystyle{naturemag}
\bibliography{references}
\end{document}